\documentclass[preprints,article,accept,moreauthors,pdftex]{Definitions/mdpi} 

\usepackage[flushleft]{threeparttable}

\firstpage{1} 
\pubvolume{1}
\issuenum{1}
\articlenumber{0}
\pubyear{2022}
\copyrightyear{2022}
\datereceived{} 
\dateaccepted{} 
\datepublished{} 
\hreflink{https://doi.org/} 
\pdfoutput=1

\Title{Multi-messenger constraints on the Hubble constant through combination of gravitational waves, gamma-ray bursts and kilonovae from neutron star mergers}

\TitleCitation{Constraints on the Hubble constant through combination of gravitational waves, gamma-ray bursts and kilonovae from neutron star mergers}

\Author{Mattia Bulla $^{1,}$*\orcidA{}, Michael W. Coughlin$^{2}$\orcidB{}, Suhail Dhawan$^{3}$\orcidC{}, and Tim Dietrich $^{4,5}$\orcidD{}}

\AuthorNames{Mattia Bulla, Michael W. Coughlin, Suhail Dhawan and Tim Dietrich}

\AuthorCitation{Bulla, M.; Coughlin, M. W.; Dhawan, S.; Dietrich, T.}

\address{%
$^{1}$ \quad The Oskar Klein Centre, Department of Astronomy, Stockholm University, AlbaNova, SE-10691 Stockholm, Sweden; mattia.bulla@astro.su.se\\
$^{2}$ \quad School of Physics and Astronomy, University of Minnesota,
Minneapolis, Minnesota 55455, USA; cough052@umn.edu\\
$^{3}$ \quad Kavli Institute for Cosmology and Institute of Astronomy, University of Cambridge, Madingley Road, Cambridge CB3 0HA, UK; sd919@cam.ac.uk\\
$^{4}$ \quad Institut f\"{u}r Physik und Astronomie, Universit\"{a}t Potsdam, Haus 28, Karl-Liebknecht-Str. 24/25, 14476, Potsdam, Germany; tim.dietrich@uni-potsdam.de\\
$^{5}$ \quad Max Planck Institute for Gravitational Physics (Albert Einstein Institute), Am M\"{u}hlenberg 1, Potsdam 14476, Germany; tim.dietrich@aei.mpg.de
}

\corres{Correspondence: mattia.bulla@astro.su.se;}

\abstract{The simultaneous detection of gravitational waves and light from the binary neutron star merger GW170817 led to independent measurements of distance and redshift, providing a direct estimate of the Hubble constant $H_0$ that does not rely on a cosmic distance ladder nor assumes a specific cosmological model. By using gravitational waves as ``standard sirens'', this approach holds promise to arbitrate the existing tension between the $H_0$ value inferred from the cosmic microwave background and those obtained from local measurements. However, the known degeneracy in the gravitational-wave analysis between distance and inclination of the source lead to a $H_0$ value from GW170817 that was not precise enough to resolve the existing tension. In this review, we summarize recent works exploiting the viewing-angle dependence of the electromagnetic signal, namely the associated short gamma-ray burst and kilonova, to constrain the system inclination and improve on $H_0$. We outline the key ingredients of the different methods, summarize the results obtained in the aftermath of GW170817 and discuss the possible systematics introduced by each of these methods. }

\begin{document}



\section{Introduction}

The Hubble constant ($H_0$) measures the present expansion rate of our Universe and sets its absolute distance scale. In the local Universe, $H_0$ can be approximated by a simple linear equation
\begin{equation}
\label{eq:H0}
    v_H = c\,z = H_0\,D_L
\end{equation}
where $D_L$ is the luminosity distance and $v_H$ is the Hubble flow velocity, equal to the speed of light $c$ times the redshift $z$. The exact value of $H_0$ has been the subject of a myriad of studies and of contentious debate from the first measurement by Edwin Hubble in 1929 \cite{Hubble:1929ig} to the present days nearly a century later. Most notably, an increasing tension has emerged between $H_0$ values measured from probes of the early Universe, e.g., the cosmic microwave background (CMB, \citep{Planck2020}), and those inferred from probes of the late Universe, e.g., Type Ia supernovae (SNe Ia, \cite{riess2021:h0}). Whether this tension is caused by hidden systematics or indicative of new physics is hotly debated (see, e.g., \cite{Verde:2019ivm} for a recent review). 

Gravitational waves (GWs) from compact object mergers have been proposed as ``standard sirens'' \cite{Schutz:1986gp,Holz:2005df} to measure $H_0$, where the source distance $D_L$ is inferred directly from the GW signal while its redshift $z$ is obtained either from an electromagnetic (EM) counterpart and its parent galaxy or statistically from galaxy clustering. This independent approach to measure $H_0$
holds promise to arbitrate the existing tension and has been vitalized following the observation of the binary neutron star (BNS) merger GW170817 on Aug 17, 2017 \cite{LIGOScientific:2017ync}, made by the Advanced Laser Interferometer Gravitational-wave Observatory (LIGO, \cite{LIGOScientific:2014pky}) and Virgo \cite{VIRGO:2014yos} detectors. The simultaneous detection of GW and EM radiation from this single source led to independent measurements of the distance and redshift of the source, thus providing a direct estimate of $H_0$ \cite{LIGOScientific:2017adf}. Nevertheless, a degeneracy in the GW signal between distance and inclination of the merging system translates into large ($\sim15\%$) uncertainties on $H_0$, with the inferred value currently unable to resolve the tension between early- and late-Universe measurements of the Hubble constant.

In this respect, an independent estimate of the system inclination can reduce the existing degeneracy and provide better constraints on $H_0$. Fortunately, not only the GW
signal but also the EM emission from neutron star (NS) merger is viewing-angle
dependent. This anisotropic emission includes a non-thermal short gamma-ray burst (GRB) ``afterglow'' powered by the interaction between a relativistic jet and the circum-burst environment, and a thermal ``macronova/kilonova'' (KN) powered by the radioactive decay of $r-$process nuclei synthesised during and after the merger. A constraint on the viewing angle from these EM probes can pin down the inclination of NS mergers,
helping relieve the degeneracy between distance and inclination and thus improving on $H_0$. Improved constraints on $H_0$ have been presented for GW170817 using constraints on the inclination from model fitting of the associated gamma-ray burst GRB\,170817A \cite{Guidorzi:2017ogy,Hotokezaka:2018dfi,Wang:2020vgr} and KN AT\,2017gfo \cite{Dhawan:2019phb,Coughlin:2020ozl,Dietrich:2020efo,Perez2022}.

This review article aims at summarizing the recent efforts to constrain $H_0$ from NS mergers using GWs alone as standard sirens or through combination of GWs and EM light. The article is organized as follows. Section~\ref{sec:h0} summarizes the current status of the $H_0$ tension, while Section~\ref{sec:sirens} outlines the key ingredients of the standard siren approach to measure $H_0$ with GWs. The rest of the article provides an overview of the various works in the literature using additional information from GRBs (Section~\ref{sec:grb}) and KNe (Section~\ref{sec:kn}) in combination with GW data to improve on $H_0$. Finally, Section~\ref{sec:discussion} provides a summary and some discussion about the systematics introduced by the different approaches.






\section{The Hubble constant tension}
\label{sec:h0}

In recent years, a remarkable increase in
accuracy obtained by a broad range of independent cosmological observations has provided compelling support for our current standard $\Lambda$ cold dark
matter ($\Lambda$CDM) model. This concordance cosmology successfully explains the measurements of fluctuations in the temperature and polarization of the CMB radiation \citep{Planck2020} as well as observations of large-scale structure and matter fluctuations
in the universe, e.g. baryon acoustic oscillations (BAO; \cite{2019MNRAS.486.2184M}).

With improved accuracy of recent observations, some discrepancies have been noted. The {\it prima facie} most significant tension, now at the 5$\sigma$ level of significance, is between the CMB inferred value of the Hubble constant, $H_0=67.4\pm0.5$ km s$^{-1}$ Mpc$^{-1}$, and the direct measurement of its local value \citep{riess2021:h0}. The local measurements are based on a calibration of the absolute luminosity of SNe~Ia using independent distances to host galaxies of nearby SNe~Ia, known as the ``cosmic distance ladder".
This claimed tension, if confirmed, could provide evidence for new fundamental physics beyond the standard model of cosmology (e.g., see \cite{Knox2020,Shah2021} for a summary of the potential non-standard cosmologies as a solution). It could, however, be a sign of unknown sources of systematic error. At the early Universe end of the distance scale, independent evaluations of $H_0$ with the Atacama Cosmology Telescope (ACT) combined with the Wilkinson Microwave Anisotropy Probe (WMAP) missions \citep[][]{Aiola2020} yield a low value of $H_0$. Moreover, replacing the CMB entirely with big bang nucleosynthesis for early universe constraints on $H_0$ show a consistency between the two methods, e.g., \cite{Addision2018,Abbott2018DES}.  
Currently, the local $H_0$ methods have slight differences in their values. The tip of the red giant branch (TRGB; \citep{2021ApJ...919...16F}) and Cepheid \citep{riess2021:h0} distance scales yield values of 69.8 $\pm$ 1.7 and 73.04 $\pm$ 1.04 km s$^{-1}$ Mpc$^{-1}$, respectively. Understanding these differences is important to discern whether the tension is a sign of novel physics or a yet-to-be-revealed systematic error. To date, only the TRGB and Cepheid measurements have measured $H_0$ at the $\sim 2 - 3 \%$ or better precision. Therefore, while the distance ladder estimates are precise, given the $\sim 2 \sigma$ level internal discordance between the different methods,  new and  independent measurements of $H_0$ at the $\sim 1-2\%$ level are imperative. 

Promising methods for precise determination of $H_0$, independent of the distance ladder include, Type II-P SNe (SNIIP; \cite{Vogl2020}), megamaser distances \citep{Pesce2020} and time-delay cosmography using lensed transients. While strongly lensed quasars have previously been shown to as precise estimators of $H_0$ \citep{Wong2020}, sources of systematics, e.g. the mass-sheet transformation need to be account for and hence, the final estimate does not, as yet, have competitive uncertainties \citep{Birrer2020}. These can be overcome with the use of lensed SNe~Ia, as standardisable candles, with future surveys like Rubin Observatory's Legacy Survey of Space and Time (VRO/LSST, \citep{Birrer2022}).  SNIIP's and megamaser galaxies are currently hampered by large peculiar velocity uncertainties since the brightness of these sources has only allowed measurements in the nearby Hubble flow, which is expected to change with future observatories. These would be highly complementary to the EM-GW constraints on $H_0$ from standard sirens, which we will discuss in Section~\ref{sec:sirens}.



\section{Gravitational waves as standard sirens}
\label{sec:sirens}

The use of GW observations to probe the expansion history of the Universe and determine $H_0$ was originally proposed by Schutz in 1986 \cite{Schutz:1986gp} and further developed by Holz \& Hughes \cite{Holz:2005df} who first introduced the term ``standard sirens''\footnote{As admitted by the authors, the term was coined by Sterl Phinney and Sean Carroll.} as opposed to ''standard candles'' to stress the aural rather than visual nature of the GW signals. This method relies on the fact that the strain amplitude $h$ of the GW is inversely proportional to the luminosity distance $D_L$ according to Einstein's quadrupole formula \cite{Einstein:1918btx} 
\begin{equation}
h = \frac{2 G}{c^4} \frac{1}{D_L}\frac{d^2I}{dt^2}\,\,\,\, ,
\end{equation}
where $G$ is the gravitational constant and $I$ is the source's mass quadrupole moment. Both $h$ and the rate of change of the frequency $df_{\rm GW}/dt$ of the signal depend on the binary masses $m_1$ and $m_2$ through the same combination, i.e.\ the so-called ''chirp'' mass $\mathcal{M}_c=(m_1\,m_2)^{3/5}\,(m_1+m_2)^{-1/5}$. The luminosity distance $D_L$ can therefore be extracted from the GW signal by measuring the two observables $h$ and $df_{\rm GW}/dt$. At the same time, the redshift of the source can be inferred directly if an EM counterpart and its host galaxy are identified (``bright sirens'') or otherwise statistically based on galaxy clustering (''dark sirens''). From the independent measurements of $D_L$ and $z$, $H_0$ is obtained directly from Equation~\ref{eq:H0} and without relying on a cosmic distance ladder or assuming a specific cosmological model (see Section~\ref{sec:h0}).

Focusing on the mergers of binary black holes (BBHs), Holz \& Hughes \cite{Holz:2005df} found that the luminosity distance $D_L$ could be measured with an accuracy of $\delta D_L /D_L\sim 1-10\,\%$ for a single event.
However, the relatively large uncertainties can be greatly reduced if an EM counterpart to the GW event is identified. First, an EM counterpart identification reduces the pointing errors and breaks the correlations in the GW signal between position and distance, therefore improving the accuracy on $D_L$ to $\delta D_L /D_L\lesssim 1\,\%$. Second, an EM counterpart leads to the identification of the host galaxy, from which a redshift can be extracted and used to directly measure $H_0$. Therefore, while the statistical ''dark siren'' method may lead to $1\%-$level accuracy on $H_0$ in the long run, e.g., \cite{MacLeod:2007jd,DelPozzo:2011vcw,Messenger:2011gi,Petiteau:2011we,Oguri:2016dgk,Chen:2017rfc,LIGOScientific:2018gmd,Gray:2019ksv,Chatterjee:2021xrm,Ghosh:2022muc}, the ''bright siren'' method remains a better prospect since the knowledge of the source redshift greatly improves the $H_0$ determination on single events \cite{Dalal:2006qt,Nissanke:2009kt,Nissanke2013,Feeney:2018mkj,Mortlock:2018azx,Feeney:2020kxk,Chen:2017rfc} (see also \cite{Trott:2021fnx} for biases in $H_0$ from the dark sirens). Nevertheless, a number of $\sim50-200$ GW events \cite{Chen:2017rfc,Feeney:2018mkj} with an identified EM counterpart might be needed to reach $1\%-$ level accuracy on $H_0$ and arbitrate the current tension (although the exact number is subject to uncertainties like, e.g., the BNS and BH-NS merger rates).

\begin{figure}[H]
\includegraphics[width=14.25cm]{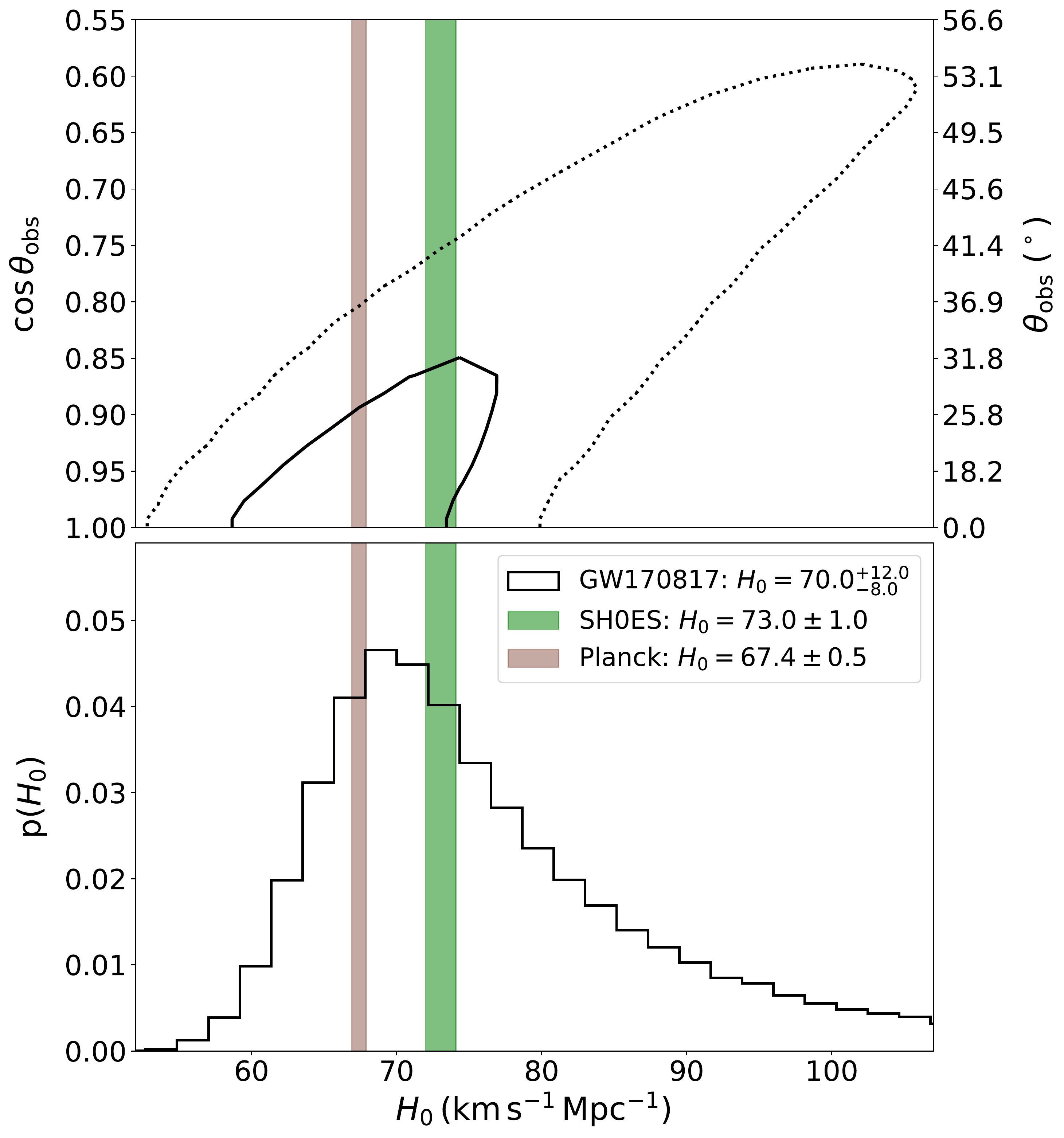}
\caption{Constraint on $H_0$ from the BNS merger GW170817 \cite{LIGOScientific:2017adf}. (Top) 2D posterior density of $H_0$ and the viewing angle $\theta_{\rm obs}$, where $68.3\,\%$ ($1\sigma$) and $95.4\,\%$ ($2\sigma$) contours are shown with solid and dotted black lines, respectively. The viewing angle $\theta_{\rm obs}$ is calculated relative to a face-on observer, i.e. $\theta_{\rm obs}=180^\circ-i$, where $i$ is the system inclination obtained from the GW data. (Bottom) Marginalized 1D posterior density for $H_0$. In both panels, $H_0$ values inferred from Planck \cite{Planck2020} and SHOES \cite{riess2021:h0} are shown with their $1\sigma$ intervals in brown and green, respectively. The inferred $H_0$ values are reported in the legend. \label{fig:H0sirens}}
\end{figure}

The power of the standard siren approach was showcased in the aftermath of GW170817 \cite{LIGOScientific:2017adf}. The detection of the short gamma-ray burst GRB170817A within a region compatible with the LIGO and Virgo sky localization for the GW source \cite{LIGOScientific:2017zic,Goldstein:2017mmi,Savchenko:2017ffs} prompted an intensive follow-up campaign \cite{LIGOScientific:2017ync} that led to the discovery of an optical transient originating from the radioactive decay of $r-$process nuclei synthesised in the merger of two NSs (a KN, see Section~\ref{sec:outflows}). This transient was located in the outskirts of the NGC\,4993 galaxy \cite{Coulter:2017wya,DES:2017kbs,Valenti:2017ngx,Arcavi:2017xiz,Tanvir:2017pws,Lipunov:2017dwd} with a chance association deemed to be at the level of $0.004\%$ \cite{LIGOScientific:2017adf}. Thanks to the EM counterpart identification, the luminosity distance of GW170817 could be constrained from the GW signal alone to $D_L=43.8^{+2.9}_{-6.9}$\,Mpc. At the same time, a redshift was inferred for NGC\,4993; after correcting for peculiar velocities, the measurement led to a Hubble flow velocity $v_H = 3017\pm166$\,km s$^{-1}$. By combining the distribution for $D_L$ from the GW signal and that for $v_H$ from the EM data, a posterior distribution on $H_0$ was inferred and a maximum a posteriori (MAP) value with $68.3\,\%$ credible interval was constrained to be $H_{0,\rm MAP}=70.0^{+12.0}_{-8.0}$ km s$^{-1}$ Mpc$^{-1}$. For completeness, we note that a $68.3\,\%$ symmetric interval (median plus the $15.85-84.15\,\%$ range) of $H_{0,\rm Symm}=74.0^{+16.0}_{-8.0}$ km s$^{-1}$ Mpc$^{-1}$ is also quoted for this measurement in the literature. As shown in Figure~\ref{fig:H0sirens}, this value is consistent with both CMB \cite{Planck2020} and local distance-ladder \cite{riess2021:h0} measurements and demonstrate the potential of the standard siren method. Owing to the $\sim15\,\%$-level uncertainties, however, the inferred $H_0$ value can not arbitrate the existing tension and multi-messenger detections of future compact binary mergers are needed to achieve this in a statistical sense \cite{Feeney:2018mkj}. 

The uncertainties on $H_0$ derived by Abbott et al. \cite{LIGOScientific:2017adf} are driven by those on the luminosity distance $D_L$\footnote{We note that another relatively large contributor to the error budget is given by uncertainties on peculiar velocities \cite{Howlett:2019mdh,Nicolaou:2019cip,Mukherjee:2019qmm}, which however are expected to be negligible at the large distances at which NS mergers will be detected in the future \cite{LIGOScientific:2014pky,Petrov:2021bqm}.}, which are largely due to a well-known degeneracy between distance $D_L$ and orbital inclination $i$ of the binary. This degeneracy can be easily seen from the following equation for $h$ valid in the limit of small angles:
\begin{equation}
h \propto \frac{\cos i}{D_L}    
\end{equation}
That is, a similar GW amplitude $h$ is produced by a binary far away and viewed close to face-on/face-off and a binary nearer by but viewed at larger inclinations. In principle, constraints on the inclination $i$ can be extracted directly from the GW data by measuring the two polarization amplitudes
\begin{equation}
h_+ \propto 2\,(1+\cos^2i) \,\,, \hspace{2cm}h_\times \propto 4\,\cos i
\end{equation}
where the constant of proportionality is the same for both $h_+$ and $h_\times$ and depends on the chirp mass $\mathcal{M}_c$, luminosity distance $D_L$ and frequency $f$ \cite{Sathyaprakash:2009xs}.  In practice, this is challenging for two main reasons.  First, only upper limits on the viewing angle $\theta_{\rm obs}$ = min[$i$,$180-i$]\footnote{Note that the viewing angle $\theta_{\rm obs}$ is measured from the jet axis while the inclination $i$ from the axis orthogonal to the binary's orbital plane. Therefore, this relation between $\theta_{\rm obs}$ and $i$ assumes that the jet axis is orthogonal to the orbital plane.} can be achieved at small inclinations since $h_+\sim h_\times$. Second, the LIGO detectors are almost aligned and thus sensitive to only one polarization, hence, a third detector (e.g. Virgo) is needed to measure both polarization amplitudes with good accuracy. Indeed, GW170817 was observed at relatively small angles and seen only by the two LIGO detectors. As shown in Figure~\ref{fig:H0sirens}, this led to poor constraints on the inclination angle of $119^\circ<i<171^\circ$ ($9^\circ<\theta_{\rm obs}<61^\circ$). Recent simulations \cite{Chen:2018omi} show that, in the absence of an EM counterpart, the International Gravitational-Wave Observatory Network (IGWN) at design sensitivity would be able to constrain the inclination of the system only if $\theta_{\rm obs}\gtrsim75^\circ$, i.e.\ for systems that are close to edge-on and thus more difficult to detect in GWs.

While it might be challenging to pin down the inclination angle from the GW data alone, constraints can be placed by exploiting the viewing-angle dependence of the EM signals. In particular, both GRB and KNe are not isotropic signals and constraining the orientation from these EM probes can alleviate the distance-inclination degeneracy in the GW analysis, thus offering a promising pathway to improve on $H_0$ on a single-event base. In the following, we will review studies using GRB (Section~\ref{sec:grb}) and KN (Section~\ref{sec:kn}) observables to constrain the system inclination and improve on $H_0$. The systematic effects introduced by these approaches will be mentioned in each section and, together with those from the GW standard siren approach, further expanded in Section~\ref{sec:discussion}. The main results in terms of inferred viewing angles and $H_0$ values are summarized in Figure~\ref{fig:fig_H0}
and Table~\ref{tab:H0}.




\section{Inclination constraints from the gamma-ray burst}

\label{sec:grb}

\subsection{Afterglow}
\label{sec:afterglow}

The association between short GRBs and NS mergers, originally put forward in the late 1980's \cite{Blinnikov1984,Goodman:1986az,Paczynski:1986px,Eichler:1989ve}, led to the conjecture that a relativistic jet could be launched from the compact object formed after the merger of BNS and BH-NS binaries. In this scenario, the interaction between the relativistic jet and the surrounding circum-merger environment is believed to produce a long-lasting non-thermal emission, the so-called ''afterglow'' powered by synchrotron radiation from electrons accelerated within the shocked medium and shining from X-ray to radio wavelengths. This paradigm was spectacularly confirmed by the detection of the short GRB\,170817A \cite{LIGOScientific:2017zic,Goldstein:2017mmi,Savchenko:2017ffs} and its afterglow \cite{Alexander:2017aly,Haggard:2017qne,Hallinan:2017woc,Margutti:2017cjl,Troja:2017nqp} in association with the GW-detected BNS merger GW170817. For more details about the theory and observations of short GRBs and their afterglow, we refer the reader to the multiple reviews in the literature, e.g., \cite{Piran:2004ba,Meszaros:2006rc,Nakar:2007yr,Gehrels:2009qy,Berger:2013jza,Kumar:2014upa}. Here we focus on the viewing-angle dependence of the afterglow signal.

Radiation from a relativistic jet moving at a velocity $\beta=v/c\sim1$ is beamed into a narrow cone with an half-opening angle $\Delta\theta\sim1/\Gamma$, where $\Gamma=(1-\beta)^{-1/2}$ is the Lorentz factor. In particular, radiation propagating from the edge of the jet is beamed into a cone around the jet half-opening angle $\theta_{\rm j}$. An ``on-axis'' observer within this narrow cone ($\theta_{\rm obs}-\theta_{\rm j}\lesssim\Delta\theta\sim1/\Gamma$) would detect the beamed radiation, while an ``off-axis'' observer outside the cone ($\theta_{\rm obs}-\theta_{\rm j}\gtrsim\Delta\theta\sim1/\Gamma$) would be blind to it. As the jet decelerates with time due to the interaction with the circum-merger environment, the $1/\Gamma$ cone widens and the afterglow emission becomes detectable at larger and larger inclinations. As a consequence, the afterglow light curve for an off-axis observer is predicted to rise continuously until the jet decelerates enough to include the observer. The gradual rise of the light curve reaches a peak at a Lorentz factor $\Gamma\sim1 / (\theta_{\rm obs}-\theta_{\rm j}$) and then start declining. Provided that the jet dominates the afterglow emission and its opening angle is much smaller than the observer angle, $\theta_{\rm j}<<\theta_{\rm obs}$, analytic relations \cite{Granot:2002za,Nakar:2002gd}
calibrated using hydrodynamical simulations \cite{Gottlieb:2019vwb} show that the time and flux of the light-curve peak can be expressed as $t_{\rm peak}\propto(\theta_{\rm obs}-\theta_{\rm j})^2$ and $F_{\nu,\rm peak}\propto\theta_{\rm obs}^{-2\,p}$, where $p>0$ is the power-law index of the electrons distribution\footnote{These relations are valid for frequencies $\nu_{\rm a},\nu_{\rm m}<\nu<\nu_{\rm c}$ (where $\nu_{\rm a}$ is the self-absorption frequency, $\nu_{\rm m}$ is the synchrotron break frequency and $\nu_{\rm c}$ is the cooling break frequency), a condition that is satisfied from X-ray to radio wavelengths as long as the density of the circum-merger environment is not much higher than the one inferred for GW170817.}. That is, the viewing-angle dependence is such that an off-axis afterglow peaks later and at a lower luminosity the farther away the observer is from the jet axis.

\begin{table}[t] 
\caption{$H_0$ values obtained for GW170817 with the standard siren approach ('GW'), together with improvements using inclination constraints from model fitting of the different EM probes: the associated GRB afterglow light curve with ('GW + GRB lc + motion') or without ('GW + GRB lc') constraints on from the jet superluminal motion; and the KN broad-band photometry ('GW + KN photometry') and spectroscopy ('GW + KN spectroscopy'). The $\Delta\sigma_{H_0}/\sigma_{H_0,\rm GW}=(\sigma_{H_0,\rm GW}-\sigma_{H_0})/\sigma_{H_0,\rm GW}$ column shows the percentage improvement in the $68.3\,\%$ ($1\sigma$) interval. $H_0$ values derived from CMB and are shown for comparison. \label{tab:H0}}
\newcolumntype{C}{>{\centering\arraybackslash}X}
\begin{tabularx}{14.25cm}{lCCC}
\toprule
\textbf{Method}	& \textbf{$H_0$} (km s$^{-1}$ Mpc$^{-1}$)	& 
\textbf{$\Delta\sigma_{H_0}/\sigma_{H_0,\rm GW}$ ($\%$)} &
\textbf{Reference}\\
\midrule
GW$^{\,\,1}$		& $70.0^{+12.0}_{-8.0}$		& / & \cite{LIGOScientific:2017adf} \\
GW$^{\,\,2}$		& $74.0^{+16.0}_{-8.0}$			& / & \cite{LIGOScientific:2017adf} \\
\midrule
GW + GRB lc$^{\,\,1}$		& $75.5^{+14.0}_{-7.3}$			& 10.7 & \cite{Guidorzi:2017ogy} \\
GW + GRB lc$^{\,\,1}$		& $69.5^{+4.3}_{-4.2}$			& 61.0 & \cite{Wang:2020vgr} \\
GW + GRB lc + motion$^{\,\,2}$		& $68.1^{+4.5}_{-4.3}$			& 63.1 & \cite{Hotokezaka:2018dfi} \\
GW + KN photometry$^{\,\,1}$		&  $72.4^{+7.9}_{-7.3}$			& 34.0 & \cite{Dhawan:2019phb}\\
GW + KN spectroscopy$^{\,\,1}$
& $69.6^{+6.3}_{-4.6}$		& 53.9 & \cite{Perez2022}\\
\midrule
Planck (CMB) &  $ 67.4\pm 0.5$ & / & \cite{Planck2020} \\
SH0ES (SNe Ia) &  $73.0\pm1.0$			& / & \cite{riess2021:h0}\\
\bottomrule
\end{tabularx}
\begin{tablenotes}
\small
    \item[1] $^1$Maximum a posteriori (MAP) interval (MAP value and smallest range enclosing $68.3\,\%$ of the posterior). $^2 68.3\,\%$ symmetric interval (median plus the $15.85-84.15\,\%$ range).
\end{tablenotes}
\end{table}

The non-thermal emission in GW170817 was consistent with an afterglow observed a moderate angle away from the jet axis. X-ray and radio light curves displayed a gradual rise starting from the first detection at 9 \cite{Troja:2017nqp} and 16 \cite{Hallinan:2017woc} days, respectively, until they reached a peak at around $\sim150$\,days. The peak was later followed by a steep decline in both X-ray and radio bands. A similar behaviour was seen in the optical although the first detection had to wait $\sim110$\,days \cite{Lyman:2018qjg} for the afterglow light to start dominating over the fading KN emission. Constraints on the viewing angle of GW170817 were inferred by different studies via model fitting of the afterglow emission both pre- and post-peak, e.g., \cite{Guidorzi:2017ogy,Lazzati:2017zsj,Hajela:2019mjy,Lamb:2018qfn,Troja:2018uns,Wu:2019rla,Ryan:2019fhz,Wang:2020vgr}. Values derived span a relatively wide range of angles, $20^\circ\lesssim\theta_{\rm obs}\lesssim35^\circ$, with typical errors that are small enough to make some of these estimates incompatible to each others. This discrepancy is due in part to the different models used but also to the existing degeneracy between $\theta_{\rm obs}$ and $\theta_{\rm j}$ (see below). Tighter constraints were extracted from fitting the light curves using additional information from the Very Long Baseline Interferometer (VLBI), see Section~\ref{sec:superl}.

Notwithstanding the uncertainties in the light curve modelling, two studies in the literature \cite{Guidorzi:2017ogy,Wang:2020vgr} used the inferred constraints on the viewing angle to reduce the distance-inclination degeneracy in the standard siren approach and therefore improve on $H_0$. Guidorzi et al. \cite{Guidorzi:2017ogy} modelled the first $\sim40$\,days of X-ray and radio data of GW170817 for two different choices of the jet opening angle, $\theta_{\rm j}=5^\circ$ and $\theta_{\rm j}=15^\circ$, and for varying viewing angles, jet parameters and densities of the circum-burst material. A uniform top-hat jet model was assumed \cite{vanEerten:2010zh,vanEerten:2011fb}. The inferred viewing-angle distribution span a relatively wide range of $25^\circ\lesssim\theta_{\rm obs}\lesssim50^\circ$, with a slight preference for a jet opening angle of $\theta_{\rm j}=15^\circ$. When adding this inclination constraint as a prior to the Hubble constant determination, MAP values of $H_0=72.5^{+8.6}_{-7.1}$ and $H_0=75.5^{+14.0}_{-7.3}$ km s$^{-1}$ Mpc$^{-1}$ were found for $\theta_{\rm j}=5^\circ$ and $\theta_{\rm j}=15^\circ$, respectively. The mild ($\sim10\,\%$) improvement on $H_0$ derived with the wider opening angle is shown with light-green curves in Figure~\ref{fig:fig_H0}. A similar analysis was later carried out by Wang \& Giannios \cite{Wang:2020vgr} modelling the radio data of GW170817 up to $\sim300$\,days with a realistic jet structure derived from general-relativistic magnetohydrodynamical simulations. The modelling of this extended dataset covering the light-curve peak led to a significant improvement on the inclination constraint compared to Guidorzi et al. \cite{Guidorzi:2017ogy}, with an inferred value of $\theta_{\rm obs}=22.3\pm0.2^\circ$. We note that this precise value for $\theta_{\rm obs}$ may be partly due to the specific model adopted displaying a rather strong viewing-angle dependence. The Hubble constant was constrained to $H_0=69.5\pm4.0$ km s$^{-1}$ Mpc$^{-1}$, with the uncertainties reduced by more than a factor of two compared to those from the GW-only analysis. The $H_0$ constraints from Wang \& Giannios \cite{Wang:2020vgr} are shown in red in Figure~\ref{fig:fig_H0}.

The uncertainties in the modelling of the GRB afterglow light curves can potentially bias the inclination estimates and introduce systematics in the $H_0$ measurement. In particular, different studies \cite{Ryan:2019fhz,Takahashi:2019otc,Beniamini:2020eza,Nakar:2020pyd,Lamb:2021use,Nativi:2021qzr} have highlighted a clear degeneracy between the observer viewing angle and the jet structure adopted. For instance, Ryan et al. \cite{Ryan:2019fhz} and Takahashi \& Yota \cite{Takahashi:2019otc} showed that for both a Gaussian and a power-law jet the ratio $\theta_{\rm obs}/\theta_{\rm j}$ is much better constrained than any of the two angles individually. This was further generalized by Nakar \& Piran \cite{Nakar:2020pyd} who demonstrated that this is a fundamental degeneracy general to all the jet structures. Furthermore, they provided a simple analytical formula to extract the $\theta_{\rm obs}/\theta_{\rm j}$ ratio from the shape of the light-curve peak (see their equation 3). The value $\theta_{\rm obs}/\theta_{\rm j}\sim5.8$ required to fit the peak of GW170817 was found to be in good agreement with the $\theta_{\rm obs}/\theta_{\rm j}$ ratio derived from the aforementioned studies fitting for the two angles individually. Moreover, Lamb et al. \cite{Lamb:2021use} showed that the inclusion/omission of a lateral spreading of the blastwave can shift $\theta_{\rm obs}$ by up to a factor of $2$. All these studies caution against using viewing-angle constraints from the light curve fitting alone, unless systematic uncertainties are properly modelled and taken into account. Alternatively, more accurate constraints on the viewing angle can be achieved if additional information is used to break the degeneracy between $\theta_{\rm obs}$ and the jet structure. One promising avenue for the latter is offered by the jet proper motion from VLBI radio images, as will be discussed in the next section.




\subsection{Superluminal motion}
\label{sec:superl}

A relativistic jet moving at a velocity close to the speed of light, $\beta\sim1$, and seen from a small angle $\theta$ can appear to move on the plane of the sky with a superluminal velocity, $\beta_{\rm app}>1$ \cite{Rees1966,Sari:1999cq}. This apparent superluminal motion is caused by the jet ''chasing'' the emitted radiation along the line-of-sight to the observer with a relativistic velocity $\beta\,c\cos\theta$. As a consequence, the difference in path travelled by two signals emitted with a time interval $\Delta t$ is reduced from $\Delta s=c\,\Delta t$ to $\Delta s^\prime=c\,\Delta t - \beta\,c\Delta t\cos\theta$. Equivalently, the two signals are detected by the off-axis observer with a reduced time separation of $\Delta t^\prime=\Delta t (1 - \beta \cos\theta)$. The apparent velocity in the plane of the sky is therefore equal to the side-way distance travelled by the jet, $\beta\,c\Delta t\sin\theta$ , divided by $\Delta t^\prime$, i.e., 
\begin{equation}
\beta_{\rm app}=\frac{\beta\Delta t\sin\theta}{\Delta t (1 - \beta \cos\theta)}=\frac{\beta\sin\theta}{1 - \beta \cos\theta}.
\end{equation}
For sufficiently small angles and large velocities, $\beta_{\rm app}$ can become larger than the speed of light, hence the motion appears superluminal. The apparent velocity reaches a maximum when $\cos\theta\sim\beta$, that is for $\beta_{\rm app}=\Gamma\beta\sim\Gamma$ and at an angle $\theta\sim\sin\theta\sim1/\Gamma$. Therefore, a measurement of $\beta_{\rm app}$ around the time of peak offers a way to constrain $\Gamma$ and the angle $\theta$, which at this time is equal to $\theta_{\rm obs}-\theta_{\rm j}$ (see Section~\ref{sec:afterglow}).

\begin{figure}[H]
\includegraphics[width=14.25cm]{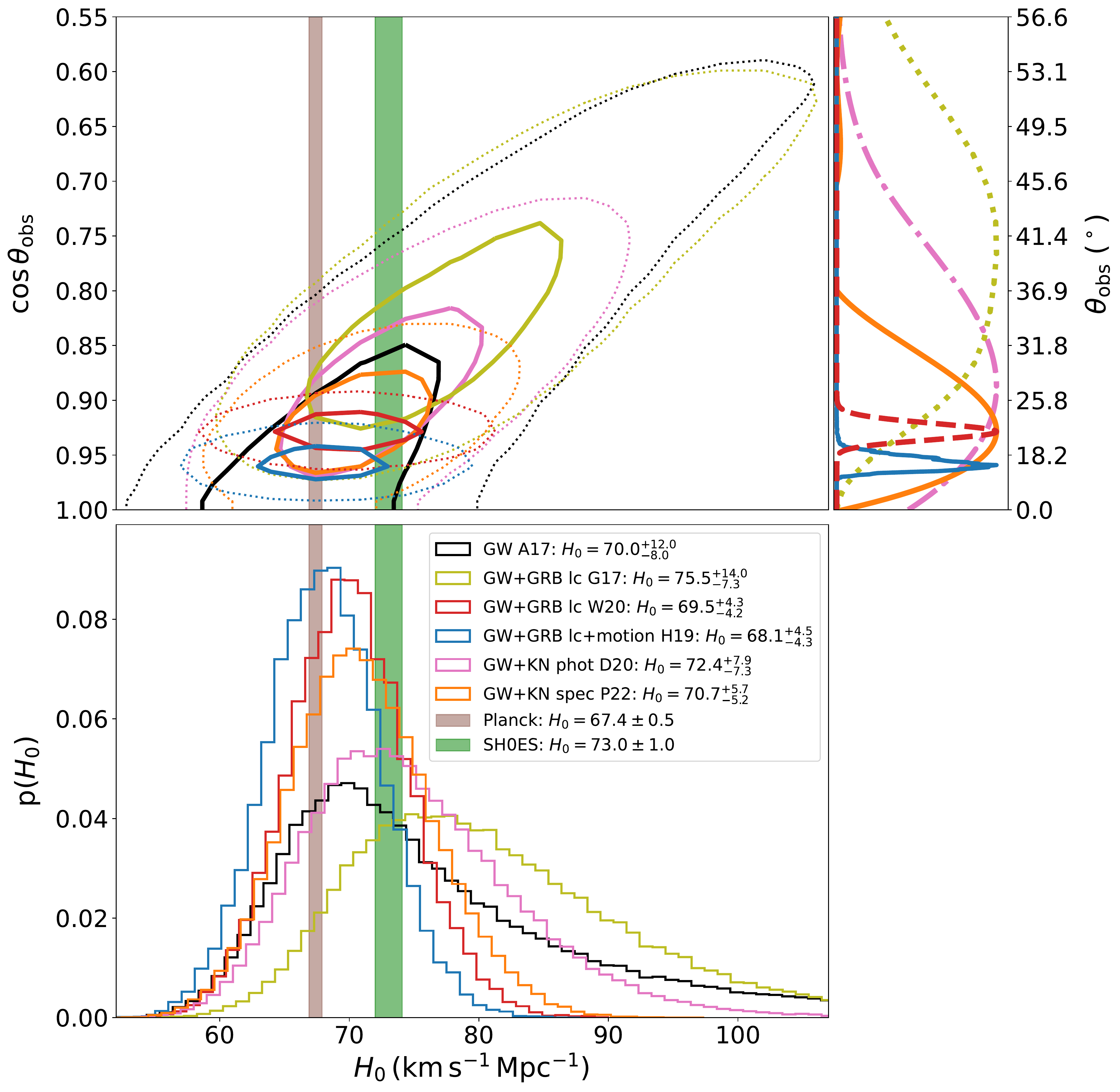}
\caption{Improved constraints on $H_0$ from the BNS merger GW170817 through combination of GW and EM data. (Top-right) Posterior distributions on the observer viewing-angle from model fitting of the associated short GRB and KN. Constraints from GRB\,170817A are shown for model fits with (cyan \cite{Hotokezaka:2018dfi}) or without (light green \cite{Guidorzi:2017ogy}; red \cite{Wang:2020vgr}) information from the jet superluminal motion \cite{Mooley:2018qfh,Ghirlanda:2018uyx}. Constraints from the KN AT\,2017gfo are shown for model fits of broad-band photometry (pink \cite{Dhawan:2019phb}) and spectroscopy (orange \cite{Perez2022}). The color scheme is the same in the remaining two panels. (Top-left) Same as in Figure~\ref{fig:H0sirens} but adding improvements to the 2D posterior density contours when the viewing-angle constraints from GRB and KN fitting are used as priors for the inclination in the GW analysis. (Bottom) Marginalized 1D posterior density distributions for $H_0$ when using the original standard siren approach (black, same as in Figure~\ref{fig:H0sirens}) and when adding constraints on the viewing-angle from EM probes. The inferred $H_0$ values are reported in the legend. \label{fig:fig_H0}}
\end{figure}

A superluminal motion of the jet was observed in the aftermath of GW170817 using radio observations with the VLBI. Mooley et al. \cite{Mooley:2018qfh} found a shift of $2.7\pm0.3$\,mas in the centroid of the radio image from 75 to 230\,days. These measurements were confirmed by Ghirlanda et al. \cite{Ghirlanda:2018uyx} with additional VLBI observations at 203\,days, finding a displacement of $2.44\pm0.32$\,mas compared to the position at 75\,days. The apparent velocity was estimated to be $\beta_{\rm app}=4.1\pm0.5$\,c, which then constrained $\Gamma\sim4$ and $(\theta_{\rm obs}-\theta_{\rm j})\sim0.25$\,rad $\sim14^\circ$ \cite{Mooley:2018qfh}. When combined with the radio light-curves of GW170817, this constraint greatly reduces the degeneracy between $\theta_{\rm obs}$ and $\theta_{\rm j}$ discussed in Section~\ref{sec:afterglow} \cite{Nakar:2020pyd} and led to more precise measurements of both angles when the afterglow light curves are fitted. Specifically, Mooley et al. \cite{Mooley:2018qfh} found that only afterglow models with a jet opening angle $\theta_{\rm j}<5^\circ$ and viewed from orientations $14^\circ\lesssim\theta_{\rm obs}\lesssim29^\circ$ are consistent with both the superluminal motion and radio light curves. The simulation providing best fit to the data corresponds to $\theta_{\rm j}=4^\circ$ and $\theta_{\rm obs}=20^\circ$. Ghirlanda et al. \cite{Ghirlanda:2018uyx} found similar values for these two angles, namely $\theta_{\rm j}=3.4\pm1^\circ$ and $\theta_{\rm obs}=15^\circ{^{+1.5}_{-1.0}}$.

The radio measurements presented in Mooley et al. \cite{Mooley:2018qfh} were used by Hotokezaka et al. \cite{Hotokezaka:2018dfi} to constrain the $H_0$ value derived from GW170817 \cite{LIGOScientific:2017adf}. While the luminosity distance was fixed to $D_L=41$\,Mpc in the previous work, this was taken as a free parameter in both the superluminal motion analysis ($14^\circ\lesssim\theta_{\rm obs}\times D_L/(41\,{\rm Mpc})\lesssim29^\circ$) and the light curve fitting. The viewing angle was constrained to $\theta_{\rm obs}\sim16.6^\circ{^{+1.7}_{-1.1}}$ or $\theta_{\rm obs}=17.2^\circ{^{+2.3}_{-2.3}}$ depending on whether a power-law or a Gaussian jet model was used. The tight constraints on the viewing angle led to a large improvements on the Hubble constant measurement, with the median and 68\% credible intervals improved from the GW-only analysis $H_{0,\rm Symm}=74.0^{+16.0}_{-8.0}$ km s$^{-1}$ Mpc$^{-1}$ to 
$H_0=68.1^{+4.5}_{-4.3}$ ($68.3^{+4.4}_{-4.3}$) km s$^{-1}$ Mpc$^{-1}$ assuming a power-law (Gaussian) jet model. The constraints derived with the power-law jet model are shown with cyan curves in Figure~\ref{fig:fig_H0}. This is a significant ($\sim60\,\%$) improvement and corresponds to a $6-7\%$ precision on $H_0$, which is however insufficient to arbitrate the Hubble constant tension. Hotokezaka et al. \cite{Hotokezaka:2018dfi} suggest that $\sim$\,15 more GW170817-like events with VLBI and light-curve data will be enough to reach a $1.8\%$ precision and potentially resolve the $H_0$ tension, in contrast to the $\sim50-100$ events required without radio data. However, Mastrogiovanni et al. \cite{Mastrogiovanni:2020ppa} carried out a detailed forecast study and argued that afterglow data (photometry + imaging) will be rare in future GW runs and may therefore not contribute significantly to narrowing down $H_0$ in the long run (see also \cite{Wang:2020vgr}).



\section{Inclination constraints from the kilonova}

\label{sec:kn}

\subsection{Matter outflows as kilonova engines}
\label{sec:outflows}

The neutron-rich outflow that is ejected during and after NS mergers is the environment from which the KN emission originates. The radioactive decay of newly synthesized, heavy elements from ''rapid neutron capture'' or ''$r$-process'' powers a thermal emission known as KN that is commonly described as a rather isotropic component potentially visible from all orientations \cite{Metzger:2019zeh,Nakar:2019fza}. The exact properties of the ejecta, e.g., their mass, velocity, geometry, and composition depend on binary properties such as the total mass, mass ratio, spin, and internal structure of the NSs. Knowledge about the ejecta properties are key to properly model KN observables and, to date, these are best revealed by numerical-relativity simulations~\cite{Hotokezaka:2012ze,Sekiguchi:2015dma,Dietrich:2016fpt,Radice:2018pdn,Kruger:2020gig,Most:2020exl,Nedora:2021eoj}.

Although potentially visible from all orientations, the KN emission is expected to be viewing-angle dependent due to the anisotropic ejection of matter in NS mergers. In general, one can distinguish different types of ejecta. 
The first class are \textit{dynamical ejecta} arising from tidal stripping of the NS(s) (tidal ejecta) and from the NS contact interface as well as core bounces (shock-driven ejecta). Tidal ejecta are distributed around the orbital plane and retain the high neutronization of the parent NS(s), i.e.\ they are generally characterized by a low-electron fraction $Y_e=n_p / (n_p+n_n)$, where $n_p$ and $n_n$ are the proton and neutron densities, respectively. In contrast, shock-driven ejecta are dominantly located in the polar regions and have an increased $Y_e$ ~\cite{Sekiguchi:2015dma,Radice:2018pdn}.  
The other main ejecta class are {\it post-merger ejecta} (also referred to as disk-wind or secular ejecta), e.g., \cite{Fernandez:2014cna,Siegel:2017nub,Fernandez:2018kax}. 
In general, post-merger ejecta are produced through remnant accretion disk winds driven by neutrino emission, magnetic fields, viscous evolution, and/or nuclear recombination energy, e.g., \cite{Just:2014fka,Radice:2018ghv,Shibata:2019wef}. Long-term hydrodynamics simulations reveal that about $\sim 10-40\%$ of the debris disk mass can get ejected, e.g., \cite{Siegel:2017nub,Radice:2018xqa}. 
The properties of the post-merger ejecta depend noticeably on the central engine: while a central magnetar will lead to ejecta with a high $Y_e$, e.g.,~\cite{Siegel:2015swa,Siegel:2015twa}, the $Y_e$ will be small if the central object is a BH formed quickly after the merger. However, in both cases the outflow is roughly spherically symmetric. In contrast, viscous effects after the merger could lead to an angular momentum transport and mass ejection centered around the orbital plane. If this happens, spiral wind outflows form and shocks in the contact region of the spiral arms can lead to an increased electron fraction of $Y_e\sim0.25$~\cite{Nedora:2019jhl}. 

The range of $Y_e$ values predicted in NS mergers leads to distinct nucleosynthesis yields across the different ejecta components. Regions of the ejecta characterized by electron fraction value $Y_e\lesssim0.25$, e.g., tidal ejecta, experience a more complete $r-$process nucleosynthesis that can reproduce the third peak observed in the solar abundances around atomic mass number $A\sim195$  \cite{Rosswog:1998hy,Freiburghaus1999}. These heavy elements include the so-called ''lanthanides'' ($140\lesssim A\lesssim175$) and ''actinides'' ($230\lesssim A\lesssim260$), hence these ejecta components are typically referred to as ''lanthanide-rich'' in the literature. Regions of the ejecta with higher values for the electron fraction, $Y_e\gtrsim0.25$, such as shock-driven and post-merger ejecta, have a small-enough ratio of free neutrons to seed nuclei that prevents the $r$-process nucelosynthesis to proceed beyond $A\gtrsim140$. These relatively high $Y_e$ values can therefore reproduce the first and second solar peak but not heavier elements such as the lanthanides, hence the corresponding ejecta component are typically referred to as ''lanthanide-free''. 

The different compositions found in the ejecta of NS mergers have profound implications in terms of the opacity of matter to radiation. While opacities of $r-$process elements are in general orders of magnitude higher than those of iron typically found in SNe \cite{Kasen:2013xka,Tanaka:2013ana}, a clear difference is seen between ''lanthanide-rich'' and ''lanthanide-free'' compositions. In the former, the multitude of line transitions from heavy elements at near-ultraviolet (UV)/optical wavelengths reprocess radiation to the infrared (IR) thus producing a so-called ''red'' KN \cite{Barnes:2013wka}. In the latter, instead, the lower opacities lead to more flux escaping at shorter wavelengths thus giving rise to a so-called ''blue'' KN \cite{Metzger:2014ila}. As we will describe in the next sections, the coexistence of multiple components within NS merger outflows with different compositions/opacities and different geometries lead to a clear viewing-angle imprint on KN observables such as light curves, spectra and polarization.

\subsection{Constraints from kilonova spectro-photometry}
\label{sec:kn_specphoto}

The KN emission in NS mergers is generally thought as a relatively isotropic component compared to the beamed GRB and its associated afterglow. As discussed in Section~\ref{sec:outflows}, however, the neutron-rich outflows ejected in these mergers comprise different components with a variety of compositions and geometry. The KN emission is therefore intrinsically anisotropic and characterized by a clear viewing-angle dependence. The viewing angle of the KN signal has been characterized in several studies either using analytic function parametrising the viewing-angle dependence \cite{Villar:2017wcc,Nicholl:2021rcr} or with detailed multi-dimensional radiative transfer simulations incorporating reprocessing effects \cite{Kasen:2014toa,Wollaeger:2017ahm,Bulla:2019muo,Darbha:2020lhz,Kawaguchi2020,Korobkin:2020spe}. The broad distinction between lanthanide-free outflows close to the jet axis and lanthanide-rich outflows around the orbital plane makes KNe increasingly fainter when moving from the jet axis (face-on view) to the orbital plane (edge-on view), an effect that can be amplified by the presence of a jet \cite{Nativi:2021qzr,Klion:2020efn}. The viewing-angle dependence of the signal varies with wavelength, time and is controlled by the detailed properties of the ejecta, but can be as strong as $\sim5$\,mag in optical bands $\sim1$ day after the merger, e.g., \cite{Darbha:2020lhz,Kawaguchi2020,Korobkin:2020spe,Perez2022}. A strong viewing-angle dependence of the signal is predicted also in terms of colors, with KNe observed at inclination closer to the jet axis being in general bluer than those observed closer to the orbital plane, e.g., \cite{Kawaguchi2020}.

The viewing-angle dependence of the KN signal has been used to place constraints on the inclination of GW170817 and therefore improve on the $H_0$ measurement from the GW standard-siren approach \cite{LIGOScientific:2017adf}. This was first done by Dhawan et al. \cite{Dhawan:2019phb}, where a two-component KN grid from Bulla et al. \cite{Bulla:2019muo} was fitted to the available near-UV/optical/IR light curves of AT\,2017gfo to extract a 1D probability distribution for the viewing angle $\cos\theta_{\rm obs}$. The result of this analysis is summarized with pink curves in Figure~\ref{fig:fig_H0}. The distribution for the viewing angle was found to peak around $\cos\theta_{\rm obs}\sim 0.87$ ($\theta_{\rm obs}\sim30^\circ$) and used as a prior for the inclination $\cos(180-\theta_{\rm obs})$ in the GW analysis. The combination of GW data and the constraint on $\theta_{\rm obs}$ led to $H_0=72.4^{+7.9}_{-7.3}$ km s$^{-1}$ Mpc$^{-1}$, which is a $34\,\%$ improvement in terms of uncertainties compared to those in Abbot et al. \cite{LIGOScientific:2017adf}. Early-time photometry was found to be crucial to pinpoint the viewing angle and reduce the uncertainties on $H_0$, with the improvement reducing to drastically when removing photometric data earlier than $\sim2$\,days.

A different approach to improve on $H_0$ was used by Coughlin et al. \cite{Coughlin:2020ozl} and applied to a sample of five KNe: the KN associated with GW170817 as in Dhawan et al \cite{Dhawan:2019phb} together with four KNe suggested \cite{Ascenzi:2018mbh} to be observed in connection with the short gamma-Ray bursts GRB 150101B27 \cite{Fong:2016irn}, GRB 05070928 \cite{Fox:2005kv}, GRB 160821B29 \cite{Kasliwal:2017qhz} and GRB 060614 \cite{Zhang:2006mb}. This effort applied KN ``standardization'' \cite{PhysRevResearch.2.022006}, where correlations between the light curve flux and color evolution, which do not depend on the overall luminosity of the transient, are compared to models. These correlations are used to obtain directly the distance modulus of each event, resulting in measurements of $H_0 = 73.8^{+6.3}_{-5.8}$\,$\mathrm{km}$ $\mathrm{s}^{-1}$ $\mathrm{Mpc}^{-1}$ and $H_0 = 71.2^{+3.2}_{-3.1}$\,$\mathrm{km}$ if models from Bulla et al. \cite{Bulla:2019muo} or Kasen et al. \cite{2017Natur.551...80K} are chosen, respectively.

While these first studies \cite{Dhawan:2019phb,Coughlin:2020ozl} derived constraints on $H_0$ using broad-band photometric data of GW170817, a recent work highlighted the key role of spectroscopy to reduce the uncertainties on $H_0$ even further \cite{Perez2022}. This study was carried out in the context of MAAT (Mirror-slicer Array for Astronomical Transients), an Integral Field Unit (IFU) to be mounted on the OSIRIS spectrograph of the 10.4\,m Gran Telescopio CANARIAS (GTC) and planned to become operational from 2023 \cite{Prada2020}. Using the Very Large Telescope (VLT) X-shooter spectra of AT\,2017gfo \cite{Smartt:2017fuw,Pian:2017gtc} as seen by MAAT, P{\'e}rez-Garc{\'\i}a et al. \cite{Perez2022} found that fitting a BNS KN grid \cite{Dietrich:2020efo} to the entire $3600-9000$\,\AA{} spectral range brings a clear improvement to the viewing-angle constraint of GW170817 compared to the one from broad-band photometry (see orange line in Figure~\ref{fig:fig_H0}, right panel). The inferred MAP value of $H_0=69.6^{+6.3}_{-4.6}$ is a $\sim54$\,\% improvement from Abbott et al. \cite{LIGOScientific:2017adf}, which is significantly higher than the 34\,\% from near-UV/optical/near-IR photometry \cite{Dhawan:2019phb} and the 5\,\% when restricting to $gri$ broad-band photometry in the same $3600-9000$\,\AA{} range observed by MAAT \cite{Perez2022}. This study highlights how critical early-time spectroscopy of KN candidates is to improve $H_0$ at a significant level from single sources.

Similarly to the case of GRB modelling (see Section~\ref{sec:afterglow}), the modelling of KNe can potentially introduce systematic uncertainties on the inclination angle and bias the $H_0$ value inferred \cite{Doctor:2019odr}. Our understanding on the physical processes controlling the KN emission is still incomplete and various source of uncertainties are expected to impact the predicted KN emission, including but not limited to the role of the ejecta geometry \cite{Heinzel:2020qlt} and $r-$process heating rates \cite{Rosswog:2016dhy}, thermalization efficiencies \cite{Barnes:2020nfi} and opacities \cite{Tanaka:2019iqp,2022arXiv220406861B}. For instance, Heinzel et al. \cite{Heinzel:2020qlt} found that the assumption about the ejecta geometry \cite{Bulla:2019muo,2017Natur.551...80K,Wollaeger:2017ahm} has a strong impact on the inclination angle, advocating for the inclusion of large systematic uncertainties ($\sim1$\,mag) when inferring the viewing angle from fits to the KN light curves. Although $H_0$ values are currently limited by statistical uncertainties, the combination of GW and KNe is expected to deliver accurate (and not just precise) $H_0$ measurements only when model systematics can be understood and controlled. Until then, these analyses should include large systematic uncertainties on the KN light curves and/or be carried out using multiple KN models, e.g., \cite{Coughlin:2020ozl,Dietrich:2020efo}.

\subsection{Constraints from kilonova polarimetry}
\label{sec:kn_pol}

As is the case for SNe \cite{Wang:2008bd}, the KN emission can be linearly polarized by Thomson scattering on free electrons that can be copious at early times when the ejected material is very hot and highly ionised. The polarization level is effectively determined by the properties and geometry of the ejected material. The ejecta properties -- such as density, temperature and composition -- determine the opacities of the ejecta and therefore how important electron scattering is compared to other processes such as bound-bound, bound-free and free-free interactions that are expected to depolarize the radiation \cite{Jeffery1991,Hoflich:1995rx,Kasen:2003ai}. The ejecta geometry, instead, control how polarizing contributions from different regions of the ejecta combine to determine the final polarization level\footnote{The polarization signal of extragalactic events as supernovae and KNe is the result of integrating over all the contributions coming from different regions of the ejecta.}: complete cancellation of all the contributions and therefore null polarization is predicted for spherically symmetric ejecta, while an incomplete cancellation and thus a net polarization is found for asymmetric ejecta. Asymmetric ejecta look differently as seen in projection from different observer orientations and the polarization signal can therefore be viewing-angle dependent. This suggests KN polarimetry as a potential probe for the inclination of the merging system.

At the relevant wavelengths (near-UV, optical, IR) and times ($\gtrsim 1$\,d) for KN emission, the dominant sources of opacity in NS mergers are bound-bound transitions from $r-$process elements and Thomson scattering \cite{Kasen:2013xka,Tanaka:2013ana,Bulla:2018rvf}. The interplay between these two processes -- the former depolarising and the latter polarising the radiation -- is therefore key to determining the final polarization state of photons escaping the system. In particular, the ratio between Thomson scattering and bound-bound line opacity, $\kappa_{\rm es}/\kappa_{\rm bb}(\lambda,t)$, is a sensitive function of wavelength and time \cite{Bulla:2018rvf}. The electron scattering opacity is wavelength independent and equal to $\kappa_{\rm es}=\sigma_{\rm Th}\times n_e / \rho $, where $\sigma_{\rm Th}=6.6524\times10^{-24}$ cm$^2$ is the Thomson cross section and $n_e$ and $\rho$ the electron and mass density, respectively. In contrast, the bound-bound line opacity from $r$-process elements shows a strong dependence on wavelength and rapidly decreases from near-UV/optical to IR wavelengths \cite{Tanaka:2019iqp}. As a result, the ratio $\kappa_{\rm es}/\kappa_{\rm bb}(\lambda,t)$ and therefore polarization state of escaping photons tend to increase moving to longer wavelengths. Moreover, the time-dependence of the $\kappa_{\rm es}/\kappa_{\rm bb}(\lambda,t)$ ratio is extremely rapid. Because of the rapid expansion and cooling of the ejecta, atoms start recombining and the number of free electrons drops. This leads to a rapid increase in bound-bound opacity and modest decrease in electron scattering opacity, i.e.\ $\kappa_{\rm es}/\kappa_{\rm bb}(\lambda,t)$ rapidly decreases with time.

\begin{figure}[H]
\includegraphics[width=14.25cm,clip=True,trim=15 270 120 30]{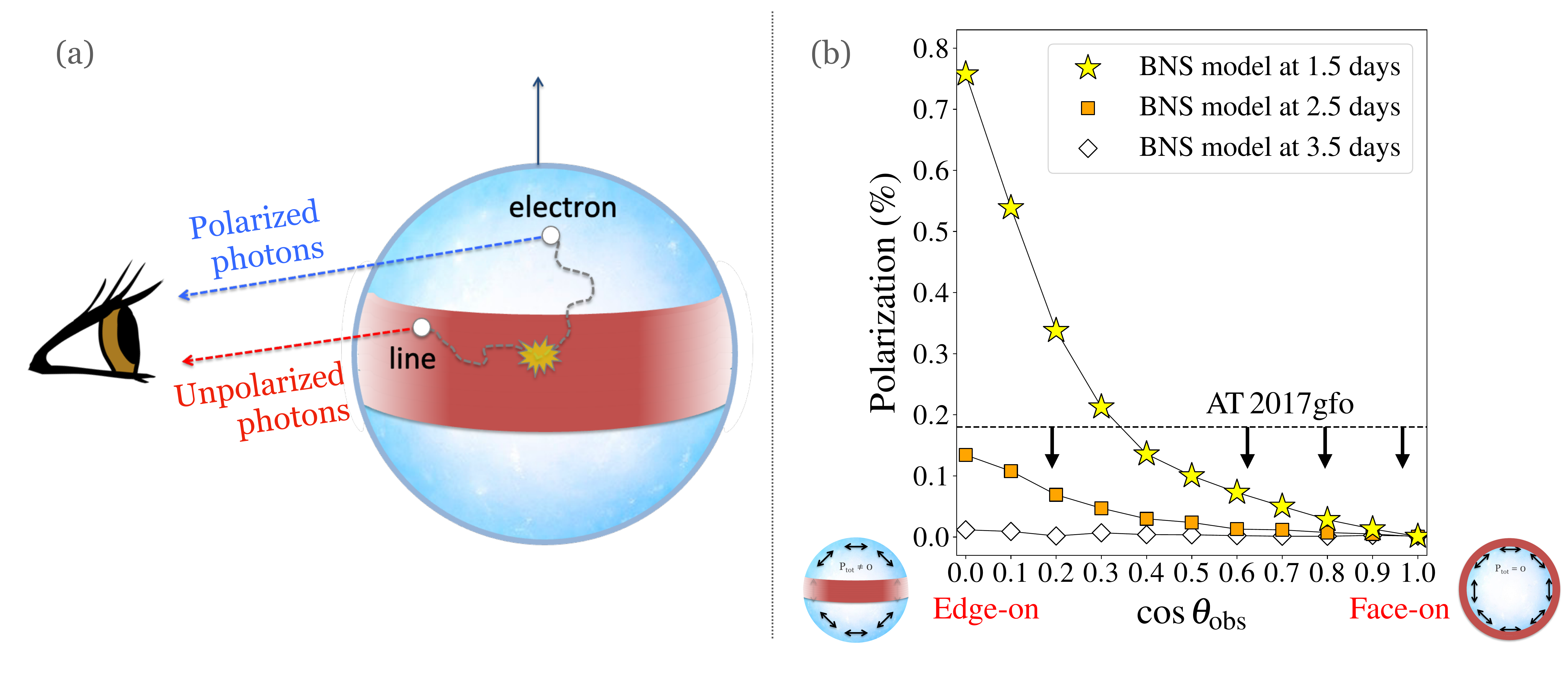}
\caption{Polarized light from KNe. (\textbf{a}) Sketch illustrating the origin of polarization in KNe. Photons escaping from lanthanide-rich dynamical ejecta around the equatorial plane (in red) are preferentially depolarized by bound-bound line interactions; photons escaping from a lanthanide-free wind (in blue) can be linearly polarized by Thomson scattering. Figure adapted from \cite{Bulla2018messenger}. (\textbf{b}) Polarization predictions from \cite{Bulla:2018rvf} for a two-component BNS model. Polarization levels are shown at 7000~\AA{} as a function of viewing angle $\theta_{\rm obs}$ for a  three different epochs: 1.5 (yellow stars), 2.5 (orange squares) and 3.5 (white diamonds) days from the merger. The $V-$band polarization upper limit derived for AT\,2017gfo at 1.5\,days is shown with a horizontal dashed line and is consistent with an observer viewing the system from an angle within $\theta_{\rm obs}\sim70^\circ$ from the jet axis ($\cos\theta_{\rm obs}\gtrsim0.35$). \label{fig:fig_pol}}
\end{figure}

The exact values of $\kappa_{\rm es}$ and $\kappa_{\rm bb}(\lambda,t)$ and therefore their relative contribution is determined by the local properties of the material ejected in NS mergers. In this respect, a key property controlling the $r$-process nucleosynthesis and thus the corresponding matter opacity is the electron fraction $Y_e$. As described in Section~\ref{sec:outflows}, at least two ejecta components with different compositions are predicted in BNS mergers: a ''lanthanide-rich'' ejecta component distributed around the orbital plane and characterized by high opacities from heavy $r$-process elements such as lanthanides and actinides; and a ''lanthanide-free'' ejecta component at higher latitudes and characterized by lower opacities from lighter $r$-process elements. As shown in Bulla et al. \cite{Bulla:2018rvf}, $\kappa_{\rm es}/\kappa_{\rm bb}(\lambda,t)<<1$ in the lanthanide-rich component from $\sim1.5$\,d onward and at wavelengths up to $\sim1.5\,\mu$m. In contrast, $\kappa_{\rm es}/\kappa_{\rm bb}(\lambda,t)\gtrsim1$ in the lanthanide-free component at optical and IR wavelengths for the first $\sim2-3$ days after the merger. As a result, radiation escaping from equatorial regions of the ejecta is typically unpolarized (except at mid-IR wavelengths and very early times) while radiation escaping from higher latitudes is polarized at optical and IR wavelengths. This effect is illustrated schematically in the left panel of Figure~\ref{fig:fig_pol}.

In BNS mergers, the combination of unpolarized light from equatorial regions and polarized light from polar regions leads to imperfect cancellation of the different polarizing contributions and thus to a net polarization signal for most viewing angles. As illustrated in Figure~\ref{fig:fig_pol}, this effect is stronger when the system is viewed edge-on (i.e.\ $90^\circ$ away from the jet axis, see ``Edge-on'' inset in the right panel) and decreases when moving towards the jet axis, at which point the ejecta becomes symmetric in projection and all the polarising contributions cancel each others (see 'Face-on' inset in the right panel). Bulla et al. \cite{Bulla:2018rvf} predict a maximum polarization level of $\sim0.8\%$ in the optical ($7000$\,\AA) 1.5~days after the BNS merger, rapidly dropping to zero within $\sim1-2$\,days (right panel of Figure~\ref{fig:fig_pol}). These predictions can be compared to polarimetric observations taken for AT\,2017gfo with the VLT FORS2 instrument \cite{Covino:2017bjc}. The $V-$band polarization of AT\,2017gfo at 1.5\,days was consistent with being caused (at least in part) by interstellar dust in our own Galaxy and an upper limit of $P_{\rm AT2017gfo}\lesssim0.18\%$ was placed by Bulla et al. \cite{Bulla:2018rvf} after careful removal of the interstellar contribution. Comparing this upper limits with the polarization predictions at 1.5\,days (Figure~\ref{fig:fig_pol}), the viewing angle of the system is constrained to be $\theta_{\rm obs}\lesssim70^\circ$ relative to the jet axis ($\cos\theta_{\rm obs}\gtrsim0.35$). 

The relatively small polarization levels expected in KNe accompanying BNS mergers and their rapid decrease with time make a polarization detection quite challenging. The upper limit on polarization derived for AT\,2017gfo translates into a poor constraint on the viewing angle, hence no attempt to improve the $H_0$ measurements was carried out by \cite{Bulla:2018rvf}. A polarization detection in future KNe may lead to a stronger constraint on the inclination angle and therefore on $H_0$. KNe accompanying BH-NS mergers might be better targets in this respect since the ejecta distribution is more strongly asymmetric compared to the BNS case. Indeed, Bulla et al. \cite{Bulla:2020jjr} found that polarization levels up to 6\,\% may be achieved in these system for favourable viewing angles and when observing at near-IR wavelengths ($\gtrsim1\mu$m). The required polarimetric accuracy at these long wavelengths ($J$, $H$ and $K$ bands) can be achieved for KNe at the distance of $150-200$\,Mpc with the instruments currently available.

Although KN polarimetry has the potential to constrain the source inclination in the future, a better exploration of the model systematics is required. In particular, the non-trivial dependence of the polarization signals on, e.g., the ejecta geometry and densities (i.e.\ masses) can bias the inferred inclination angle if the full parameter space is not sampled properly. While Bulla et al. \cite{Bulla:2018rvf,Bulla:2020jjr} focused on polarization predictions for individual models, follow-up studies should explore the range of polarization signals expected for a large grid of models varying parameters such as the geometry and masses, in a similar fashion to what was done in terms of light curves by, e.g., Dietrich et al. \cite{Dietrich:2020efo} and Anand et al. \cite{Anand:2020eyg}.



\section{Summary and outlook}
\label{sec:discussion}

The use of GWs as standard sirens \cite{Schutz:1986gp,Holz:2005df} holds promise to arbitrate the existing tension between early- \cite{Planck2020} and late- \cite{riess2021:h0} Universe probes of the Hubble constant $H_0$. Without additional information, however, this is unlikely to happen in the near future since $\sim50-200$ GW events \cite{Chen:2017rfc,Feeney:2018mkj} with an EM counterpart are needed to reach a precision of $\sim1 \%$ on $H_0$. This goal might be achieved earlier and with fewer events if a constrain on the system inclination can be obtained from the GRB and/or the KN accompanying NS mergers detected with GWs. Indeed, a few proof-of-concept studies have shown that a constrain on the viewing angle of GW170817 from the GRB afterglow \cite{Guidorzi:2017ogy,Wang:2020vgr}, the GRB superluminal motion \cite{Hotokezaka:2018dfi} and the KN \cite{Dhawan:2019phb,Coughlin:2020ozl} can greatly reduce the degeneracy between distance and inclination in the GW data and therefore the uncertainties on $H_0$. Improved constraints leading to $\sim6\,\%$-level uncertainties on $H_0$ for GW170817 can be reached when information from the different EM probes are combined together in a truly multi-messenger analysis, as shown in Dietrich et al. \cite{Dietrich:2020efo}.

Nevertheless, a few questions remain. While the required precision on $H_0$ might be achieved in the future as more NS mergers are detected in both gravitational and electromagnetic waves, the accuracy of the inferred value will depend crucially on whether the systematics from the standard siren approach are under control, as well as those from the GRB/KN modelling. First, the standard siren approach can be affected by systematic uncertainties both from the GW and the EM analysis. On the GW side, calibration errors potentially affecting the inferred $D_L$ and hence $H_0$ are subdominant compared to statistical uncertainties on single events, e.g., \cite{Vitale:2011wu,Payne:2020myg,Vitale:2020gvb}, and even when combining multiple ($\gtrsim100$) sources in the most realistic case studied by Huang et al. \cite{Huang2022}. On the EM side, possible source of systematic uncertainties come from estimates of peculiar velocities in nearby events \cite{Howlett:2019mdh,Nicolaou:2019cip,Mukherjee:2019qmm} and viewing-angle selection effects biasing discovery of EM counterparts towards systems viewed close to face-on \cite{Chen:2020dyt}. Secondly, additional source of systematics are introduced by the uncertainties on models used to fit GRB \cite{Ryan:2019fhz,Takahashi:2019otc,Beniamini:2020eza,Nakar:2020pyd,Lamb:2021use,Nativi:2021qzr} and KN \cite{Doctor:2019odr,Rosswog:2016dhy,Heinzel:2020qlt,Barnes:2020nfi} observables to constrain the inclination. In a recent study, Chen et al. \cite{Chen:2020dyt} showed that the viewing-angle effects are likely to dominate the systematics budget and be a major challenge to resolve the $H_0$ tension with GWs and light from NS mergers. Specifically, 
a $\sim2\,\%$ bias on $H_0$ can be introduced by viewing-angle selection effects and a $\sim3\,\%$ bias if the inclination constraints from EM probes are not controlled under $\sim10^\circ$.

Looking ahead, the detection of more NS mergers in both GW and light will be essential to enable a thorough understanding of the source of stystematic uncertainties and thus for the standard siren approach to show its full potential. In the next decade, the number of GW detectors and their sensitivities are expected to increase \cite{KAGRA:2013rdx} and this will be accompanied by the advent of new optical sky surveys like, e.g., the VRO/LSST \cite{LSST:2008ijt}. This combined network will facilitate multi-messenger detections of NS mergers beyond GW170817 and potentially lead to both precise and accurate values of the Hubble constant.






\vspace{6pt} 




\funding{M.B. acknowledges support from the Swedish Research Council (Reg. no. 2020-03330). M.W.C. acknowledges support from the National Science Foundation with grant numbers PHY-2010970 and OAC-2117997. S.D. acknowledges support from the Marie Curie Individual Fellowship under grant ID 890695 and a junior research fellowship at Lucy Cavendish College. T.D. acknowledges financial support through the Max Planck Society.}

\conflictsofinterest{The authors declare no conflict of interest.} 



\abbreviations{Abbreviations}{
The following abbreviations are used in this manuscript:\\

\noindent 
\begin{tabular}{@{}ll}
ACT & Atacama Cosmology Telescope \\
BAO & Baryon Acoustic Oscillations \\
BBH & Binary Black Hole \\
BH & Black Hole \\
BNS & Binary Neutron Star \\
CMB & Cosmic Microwave Background \\
GRB & Gamma-ray burst \\
GTC & Gran Telescopio CANARIAS \\
GW & Gravitational Wave \\
IFU & Integral Field Unit \\
IGWN & International Gravitational-Wave Observatory Network \\
IR & Infrared \\
KN & Kilonova \\ 
$\Lambda$CDM & $\Lambda$ Cold Dark Matter \\
LIGO & Laser Interferometer Gravitational-wave Observatory \\
LSST & Legacy Survey of Space and Time \\
MAAT & Mirror-slicer Array for Astronomical Transients \\
MAP & Maximum a posteriori \\
NS & Neutron Star \\
SNe & Supernovae \\ 
TRGB & Tip of the red giant branch \\
UV & Ultraviolet \\
VLBI & Very Long Baseline Interferometer \\
VLT & Very Large Telescope \\
VRO & Vera Rubin Observatory \\
WMAP & Wilkinson Microwave Anisotropy Probe\\
\end{tabular}}


\end{paracol}



\reftitle{References}


\bibliography{biblio.bib}


%



\end{document}